\documentstyle[twoside, epsf]{article}

\input ibvst.sty
\input ibvs2.sty

\begin{document}

\IBVShead{5553}{6 August 2004}

\IBVStitletl{HD~52452: New BVRI Photometry }

\IBVSauth{Barway, Sudhanshu; Pandey, S. K.}
\IBVSinsto{School of Studies in Physics, Pt. Ravishankar Shukla University, Raipur - 492010, India}

\SIMBADobjAlias{V369 Gem}{HD 52452}
\IBVSkey{Photometry}

\begintext

\vspace*{-0.8cm}

HD~52452 (V369~Gem, SAO~78998; $V$ = 8.05, $B-V$ = 0.67; G5 V) is   one  
of the  shortest  period (0\fday42304) non-eclipsing chromospherically 
active binary star discovered so far. Its  X-ray properties were discovered
in the  ROSAT all sky survey  program and it  is one  among 383 relatively 
bright  X-ray sources cataloged by Pounds et al. (1993). The star  HD~52452
is an optical counterpart of the EUV bright source RE J70222+255054 
(Mason et al. 1995) and its photometric and spectroscopic  observations 
aimed at the classification of EUV stellar sources detected by EXOSAT 
and ROSAT were carried by Cutispoto et al. (1999, 2000). Furthermore,  
HD~52452  is also  listed in the 2RE source  catalogue  published by 
Pye et al.  (1995). Messina et al. (2001)   ~reported that HD~52452  is a  
triple system consisting of a tidally coupled G4 V + late-G SB1 
(responsible for the most of the observed optical variability) and a 
G5 V companion. Their photometric observations reveal that the observed 
photometric variability is due to the presence of cool spots on the 
photospheres of both component of the SB1 system. In this paper we report
the multi-band BVRI photometry of HD~52452.

The  BVRI photoelectric  photometric  observations of  HD~52452 were carried
out during two observing runs - one during February 17 - March 23, 2000
 for 8 nights and another one during February 20 - February 26, 2001 for 6 
nights. For the first observing run we have a total of 158 data points and have
 93 data points for the next one. The 40-cm Schmidt-Cassegrain
 LX 200 Meade telescope equipped with SSP-3A photoelectric photometer and 
Johnson standard broad-band BVRI filters were used for the observation. 
The telescope is situated on the campus of the Inter University Centre 
for Astronomy and Astrophysics (IUCAA) in Pune, India. The detector used
 in the SSP-3A photometer is a silicon PN-photodiode which is  not cooled. 
The response function of the B, V, R and I filters with the detector 
closely match those of the Johnson standard filters. In order to  obtain 
accurate differential photometry, we used two nearby stars HD~52071 
(K2 III, $V = 7.11$, $B-V = 1.27$) as comparison star and HD~50692 
(G0 V, $V = 5.76$, $B-V = 0.56$) as check star. The mean of four to five
 independent differential magnitudes per observation in each band were 
corrected for atmospheric extinction and transformed into BVRI standard 
system. No significant light variation was detected for  the differential
 magnitudes of the comparison and check star $\Delta V_c$, which is a good
 measure of the quality of our observation. The uncertainties in $\Delta V$,
 $\Delta (B-V)$, $\Delta (V-R)$ and $\Delta (V-I)$ are 0.015, 0.02, 0.017 and 
0.02 magnitudes, respectively.  
 
\begin{table*}
\caption{Photometric periods for HD~52452}
\begin{center}
\begin{tabular}{lcc}
\hline
Data set & Period & FAP (\%) \\
\hline 
2000+Messina et al. (2001) & 0.42402$\pm$0.00021 &  2.00 \\
2001+Messina et al. (2001) & 0.45163$\pm$0.00146 &  4.40 \\
Combined (2000+2001)      & 0.42261$\pm$0.00002 &  5.68 \\
\hline
\end{tabular}
\end{center}
\end{table*}

The combined data for both 2000 and 2001 were analyzed to obtain the photometric
period using a Scargle-Press period search routine (Scargle 1982, Horne \&
Baliunas 1986), and a photometric period P =  0\fday42261$\pm$0.00002, with a false-
alarm-probability \hbox{FAP = 5.68 \%} was found. 
Photometric periods with error and false-alarm-
probability (FAP in \%) 
are listed in Table 1. To cross check our period search routine we also 
derived photometric period from Messina et al. (2001) ~data and a photometric period 
0\fday42309$\pm$0.00017 with FAP 8.61 \% was found which agrees well
with the period (0\fday42304) reported by Messina et al. (2001). The slightly larger period for the epoch 
2001 with Messina et al. (2001) ~data as compared to the other epoch may be due to the scatter
in the data covering long period. 

Observed data points have been folded in to a phase using the photometric 
ephemeris ${\rm HJD} = 2449672.0 + 0\fday42261\times E$. We divided our 
photometric data points into four subsets (two for each epoch) and phase 
diagram for these subsets  and associated colors $(B-V), (V-R)$ and $(V-I)$ are
shown in Figure 1 and 2 for the V band. The differential magnitudes 
of the check star with respect to the comparison star V$_c$ are shown in the 
same diagrams. We have also shown the  complete light curve for epoch 2000 and 
2001 in the panel (c) of Figure 1 and 2 respectively.  Table 2 gives the epoch,
JD interval, the minimum phase and maximum amplitude for observed data 
points.  

\begin{table*}
\caption{Results form photometric analysis of HD~52452}
\begin{center}
\begin{tabular}{lcccc}
\hline
Epoch & JD  & Minimum & phase  & Maximum \\
      & interval & I &  II & amplitude \\
\hline                                    
2000.132 & 51592.144 - 51593.347 & 0.2 & 0.7 & 0.13 \\
2000.193 & 51605.125 - 51608.304 & 0.2 & 0.7 & 0.24 \\
\hline
2001.140 & 51961.140 - 51962.301 & 0.1 &  0.6   & 0.14 \\
2001.147 & 51963.141 - 51964.211 & 0.1 &  0.6   & 0.16 \\ 
\hline
\end{tabular}
\end{center}
\end{table*}

A  significant variation in the light curve within the first year and between 
the two seasons from the phase diagrams is seen (Fig. 1 \& 2). Small variations 
are observed for the
epoch 2000.132 but the data points are very scattered. Visual inspection
of light curve for the star HD~52452 reveals the 
existence of two minima separated from each other by about half period in phase. 
This
behavior is possibly due to the existence of two spots on the
stellar surface. The amplitude of light variation 
is $\sim$0.24 magnitude in $V$ band for 2000.193, which is large compared to
the $\sim$0.16 magnitude in $V$ band reported by Messina et al. (2001). He also observed peaked light curve separated by 0.4 in phase. For the next 
observing epochs 2001.140 and 2001.147 the amplitude of light variation 
decreases to $\sim$0.14 and  $\sim$0.16 magnitude in $V$ band respectively. 

The 
$(V-I)$ color curve for the epoch 2000.193 reveal the variation of $\sim$0.7 
magnitude but except this, in general there are no significant variation for the $(B-V), (V-R)$ and 
$(V-I)$  color indices. The observed variation  in $(V-I)$ color 
index can be attributed to the lower spot temperature relative to the 
photosphere than in other cases. The present observations clearly suggests 
that the optical variability in the star HD~52452 is due to 
the presence of cool spots on the stellar surface. 

The comparison of our data with previous
 observations reported by  Messina et al. (2001) shows that there is a variation 
 in amplitude, but the phases of the two minima, thus the positions of the spot, are quite stable during our observations.
The very short photometric period makes this star very interesting for studying the
 evolution of spots on stellar surface in terms of spot parameters over a
 longer time period.\\

\IBVSfig{13cm}{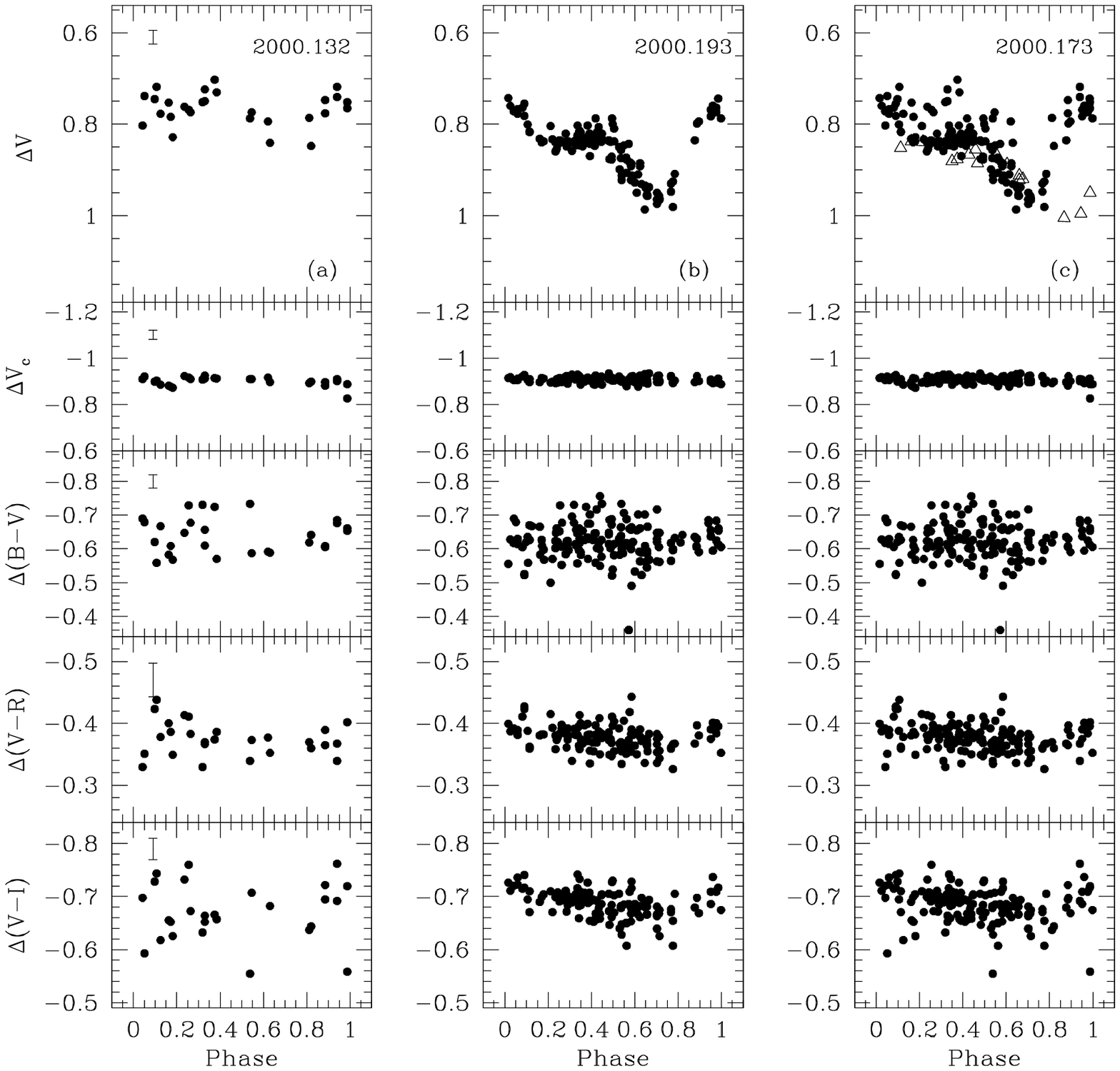}{$V$ band light curves and $(B-V), (V-R)$ 
and $(V-I)$ colors for HD~52452 for the observing run during 
February - March 2000. $V_c$ for the check star observed on the same nights. 
Typical error bars are shown in the upper left corner of each light curve in 
panel (a). The open triangle denote the observations from Messina et al. (2001).}
\IBVSfigKey{5553-f1.eps}{HD 52452}{light curve}

\IBVSTack{We are thankful to IUCAA for providing their observing, library 
and computing facilities. We also express our sincere thanks to Professor 
A. K. Kembhavi and Padmakar Singh Parihar for there kind co-operation and 
suggestions during the course of the observations. We are grateful to Dr. 
Katalin Olah whose comments and suggestions have led to significant 
improvements in the presentation of the paper. SB and SKP would like 
to express their sincere thanks to CSIR, New Delhi, for financial support 
through a project grant No. 03(0985)/03/EMR-II.}

\clearpage

\IBVSfig{13cm}{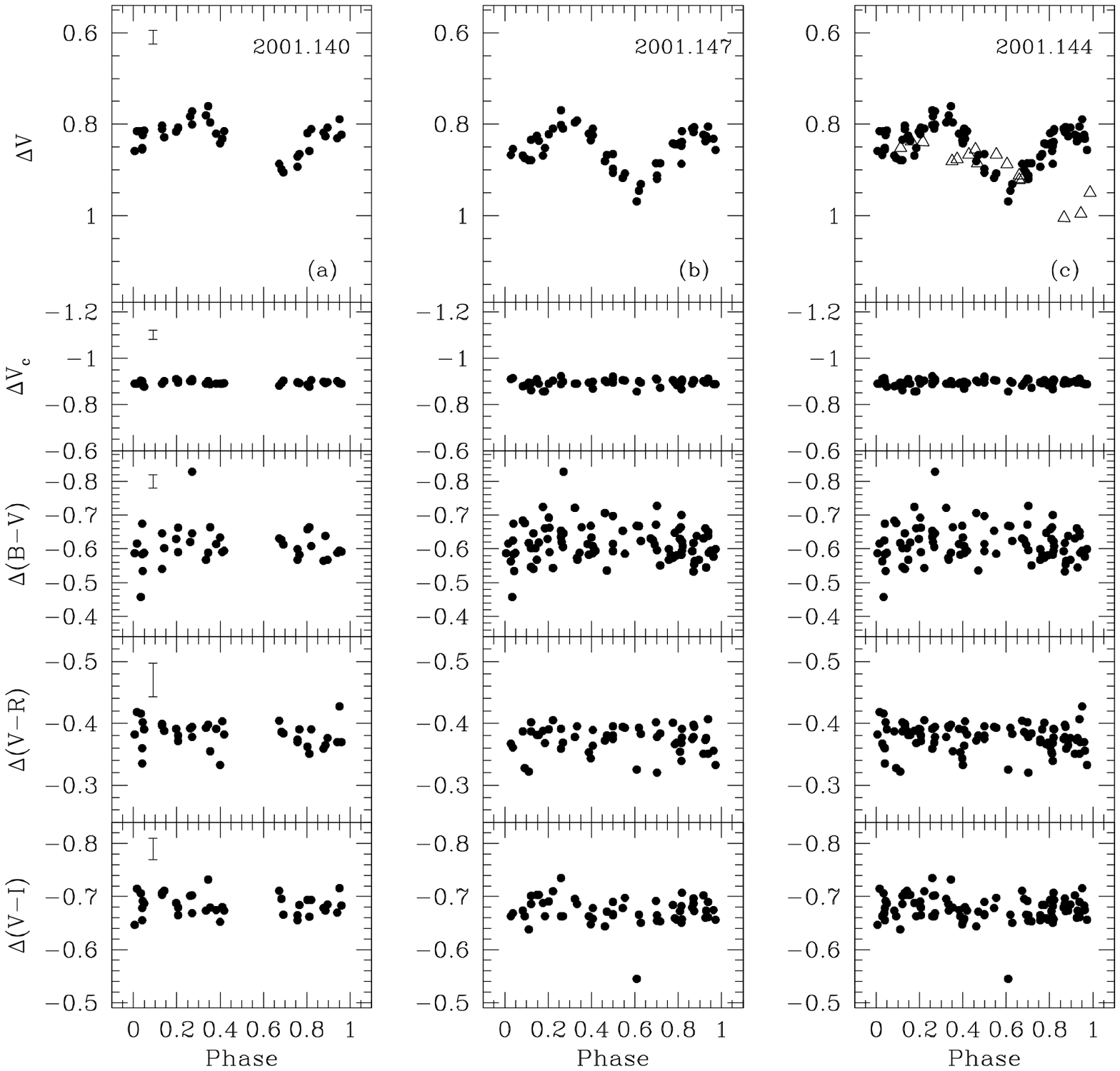}{$V$ band light curves and $(B-V), (V-R)$ 
and $(V-I)$ colors for HD~52452 for the observing run during 
February 2001. V$_c$ for the check star observed on the same nights. 
Typical error bars are shown in the upper left corner of each light curve 
in panel (a). The open triangle denote the observations from Messina et al. (2001).}
\IBVSfigKey{5553-f2.eps}{HD 52452}{light curve}

\references

Cutispoto, G., Pastori, L., Tagliaferri, G., et al., 1999, {\it A\&AS}, {\bf 138}, 87

Cutispoto, G., Pastori, L., Guerrero, A., et al.,  2000, {\it A\&A}, {\bf 364}, 205

Horne, J. H., Baliunas, S. L., 1986, {\it ApJ}, {\bf 302}, 757

Mason, K. O., Hassall, B. J. M., Bromage, G. E., et al., 1995, {\it MNRAS}, {\bf 274}, 1194

Messina, S., Cutispoto, G., Pastori, L., et al., 2001, {\it IBVS}, No. 5014

Pounds, K. A., Allan, D. J., Barber, C., et al., 1993, {\it MNRAS}, {\bf 260}, 77

Pye, J. P., Mcgale, P. A., Allan, D. J., 1995, {\it MNRAS} {\bf 274}, 1165

Scargle, J. D., 1982, {\it ApJ} {\bf 263}, 835

\endreferences

\end{document}